\documentstyle[aps,prl,multicol,epsf]{revtex}
\begin{document}

\title{Using Nonlinear Response to Estimate the Strength of an Elastic Network}

\author{Gemunu H. Gunaratne}

\address{Department of Physics, University of Houston, Houston, TX 77204}
\address{The Institute of Fundamental Studies, Kandy 20000, Sri Lanka}
\maketitle
\nobreak

\begin{abstract}
Disordered networks of fragile elastic elements have been proposed as a model
of inner porous regions of large bones [Gunaratne et.al., cond-mat/0009221, http://xyz.lanl.gov].
It is shown that the ratio $\Gamma$ of responses of such a network to 
static and periodic strain can be used to estimate its ultimate (or breaking) stress. 
Since bone fracture in older adults results from the weakening of porous bone, 
we discuss the possibility of using $\Gamma$ as a non-invasive diagnostic 
of osteoporotic bone.
\end{abstract}

\nobreak
\begin{multicols}{2}

Osteoporosis is a  major socio-economic problem in an aging 
population~\cite{kibAsmi}.  Unfortunately, therapeutic agents which can 
prevent and even reverse  osteoporosis often induce adverse side 
effects~\cite{wein}. Thus, non-invasive diagnostic tools to determine 
the necessity of  therapeutic intervention are essential for effective management 
of osteoporosis. Bone Mineral Density (BMD), or the effective bone density  
is the principal such investigative tool~\cite{cann}.  Ultrasound transmission through 
bone~\cite{njeAhan} and geometrical characteristics of the inner porous 
region or trabecular architecture (TA)~\cite{potAben,legAcha,majAnew} are being 
studied as complementary diagnostics.

In older adults, weakening of the TA is the principal cause of  increased
propensity for bone fracture~\cite{njeAhan}.  Analysis of models
can complement mechanical studies of bone in aiding the identification of
precursors of the weakening of a TA. In Ref. \cite{gunAbas}, 
it was proposed that a system to model mechanical properties of a TA 
can be obtained by adapting a disordered network of fragile 
elastic elements~\cite{chuAroo}.
The model system includes potential energy contributions from elasticity and 
from changes in bond angles between adjacent springs. 
Furthermore, springs that are strained beyond (a predetermined value) $\epsilon$ 
and bond angles that change more than $\delta$ are assumed to
fracture and are removed from the network. The inertia of the 
network is modeled by placing  masses at the vertices. When 
in motion, each mass experiences a dissipative force proportional to its 
speed. Osteoporosis is modeled by random removal of a fraction $\nu$
of springs from the network, and the bone density is estimated by the 
fraction of remaining links. Therapeutic regeneration is 
introduced by strengthening springs that experience large strain 
(as suggested by Wolff's law~\cite{cowAsad}). 

Numerical studies of the system show analogs of several mechanical properties 
of bone including, (1) an initially linear  stress vs. strain curve that 
becomes nonlinear beyond the fracture of elastic elements, 
(2) an exponential reduction of the 
ultimate (or breaking) stress with decreasing BMD, and (3) a dramatic 
increase of bone strength  following therapeutic regeneration~\cite{gunAbas}. 
Together they support the conjecture that elastic networks 
are a suitable model of mechanical properties of bone. 
In this Letter we use results from a numerical study of the model 
to introduce a possible diagnostic tool for osteoporosis.

\begin{figure}
\epsfxsize=2.00truein
\hskip 0.50truein
\epsffile{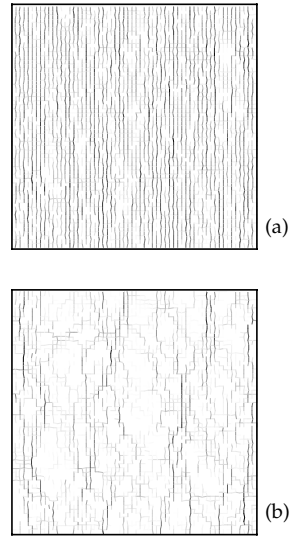}
\vskip 0.4in
\caption{The stress distributions on networks of size $60\times 60$ with  
(a) $\nu=10\%$, and (b) $\nu=30\%$ representing ``healthy" and 
``osteoporotic" bone respectively. For clarity only the compressed bonds 
are shown, and darker hues represent larger stresses. Notice that the 
``stress backbone" of (a) is dense, while that of (b) consists of a few
coherent pathways.}
\label{latt}
\end{figure}

The primary difference between  ``healthy" and ``weak" networks is the
nature of stress propagation through them.
Figure \ref{latt} provides a representation of stresses 
in two networks subjected to uniform compression~\cite{para}. 
Figure \ref{latt}(a) shows the effect on a ``healthy" network ($\nu=10\%$) 
where large stresses supporting propagation are seen to form 
a dense subset. In contrast, elastic elements supporting a 
``weak" network ($\nu=30\%$)  form a few coherent pathways (Figure \ref{latt}(b)). 
We refer to the set active in stress propagation as the ``stress 
backbone" of a network~\cite{mouAdux}. For a wide range of control 
parameters, it is seen to become finer with increasing $\nu$. 

It is easy to understand how the nature of the stress backbone is 
related to the stability of a bone. Loss of connectivity of a healthy 
TA (due to trauma) will have little effect on its stability, since many 
alternative pathways are available for stress propagation. In contrast,
fracture of a critical link (i.e., one belonging to the stress backbone) in a weak
network will have a significant impact on the remaining stress pathways, likely inducing
further fracture of elastic elements. Below, we discuss how these
variations in stress backbones effect the nonlinear response of networks
to externally applied strain.

Consider an elastic network from which a fraction $\nu$ of elastic
elements have been removed. It is first subjected to an adiabatically 
reached compression $\zeta_0$~\cite{foot2}. $\zeta_0$ is chosen well 
below the yield point so that there is no fracture of elastic elements. 
Under these conditions, the stress $T_0$ needed to maintain the compression is
proportional to $\zeta_0$~\cite{fung,gunAbas}, and the {\it static 
susceptibility} of the network is defined as $\chi_0 = T_0 / \zeta_0$.

\begin{figure}
\epsfxsize=2.50truein
\hskip 0.25truein
\epsffile{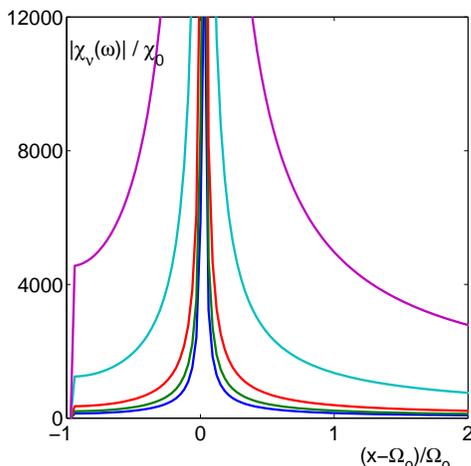}
\caption{The power spectra of the dynamical susceptibility (with 
$\zeta_d=0.001$ and $\Omega_0=100$) of several 
elastic networks normalized by the static susceptibility. The curves 
correspond to values of $\nu$ of 0\% (blue), 10\% (green), 20\% (red),
30\% (aquamarine), and 40\% (magenta).}
\label{spectra}
\end{figure}

Next, this compressed network is subjected to an additional periodic
strain $\zeta(t) = \zeta_p \exp(i\Omega_0 t)$, and
the force required for its implementation is denoted $T_{\nu}(t)$. 
$\zeta_p$ is chosen to be small ($\zeta_p \ll \zeta_0$), and hence 
$T_{\nu}(t)$ can be assumed to be proportional to
$\zeta_p$. The {\it dynamical susceptibility} of the network is defined 
by $T_{\nu}(t) = \chi_{\nu}(t)\cdot\zeta(t)$. The Fourier 
transform of $T_{\nu}(t)$ is given by the convolution 
$\hat T_{\nu}(\omega) = \hat\chi_{\nu}(\omega)*\hat\zeta(\omega)$.
Since $\hat\zeta(\omega) = \zeta_d \delta(\omega+\Omega_0)$, it follows that 
$\hat\chi_{\nu}(\omega) = \hat T_{\nu} (\omega-\Omega_0)/\zeta_d$~\cite{foot3}.

Figure \ref{spectra} shows the power spectra 
$|\hat\chi_{\nu} (\omega)| / \chi_0$ for values of $\nu$ of 0\%, 10\%, 20\%,
30\%, and 40\%.  Though both the static and
dynamical susceptibilities reduce with advancing ``osteoporosis" (increasing 
$\nu$), $\chi_{\nu} (t)$ experiences a smaller reduction. 
This is possibly due to the formation of additional (temporary) stress pathways 
when  the network is subjected to periodic strain.

\begin{figure}
\epsfxsize=2.50truein
\hskip 0.25truein
\epsffile{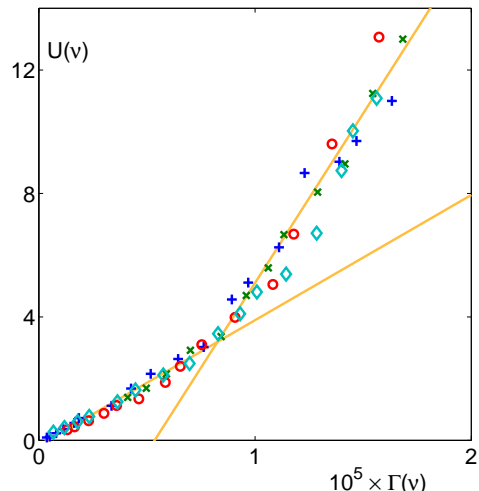}
\caption{The relationship between the ultimate (or breaking) stress $U(\nu)$
and $\Gamma(\nu)$ for several elastic networks subjected to different 
compressions $\zeta_0$. The symbols `+' (blue) and  `x' (green) represent
distinct networks with
identical control parameters compressed by  $\zeta_0=4.0$ and  $\zeta_0=8.0$ 
respectively. The symbols 'o' (red) and `$\diamond$' (aquamarine) 
represent two other network generated with different sets of control parameters,
and compressed by $\zeta_0=5.0$ and $\zeta_0=3.0$ respectively.
For these networks the values of $U(\nu)$ have been rescaled as described 
in the text.}
\label{fit}
\end{figure}

Variations of these power spectra can be quantified using the ratio 
$\Gamma (\nu)$ defined by
\begin{equation}
   \Gamma^{-1}(\nu) = \frac{1}{\chi_0} \int |\hat \chi_{\nu} (\omega)| d\omega,
\end{equation}
the range of the integral being a Nyquist frequency domain~\cite{preAfla}.
Figure \ref{fit} shows a remarkable relationship between the 
ultimate stress $U(\nu)$ and $\Gamma(\nu)$ for several networks.
The symbols `+' and `x' represent two distinct networks (constructed
using different random seeds) with identical control 
parameters~\cite{para} which have been subjected to compressions of
$\zeta_0 = 4.0$ and $\zeta_0 = 8.0$ units respectively. $U(\Gamma)$ is seen
lie on a common bi-linear curve.

The behavior of $U(\Gamma)$ for a third network is included to determine
the effects of increasing the range of elastic moduli of
the network~\cite{para2}. The ultimate stress of a network is expected to reduce
when the range of spring constants is increased (say, by a factor $f$). This is because
of the increase of the number of weak springs in the network. To compensate for
this decrease, we rescale $U(\nu)$ (heuristically) by$f$. Then,  symbols `o' representing 
the third network  fall on the curve determined by the first pair of networks.
$U(\Gamma)$ will, of course, depend  linearly on the mean elastic modulus,
as has been confirmed.

Finally, we include results from a fourth network to study variations of
the fracture strain of elastic elements. A {\it fixed-strain} fracture criterion was used in
the model; i.e., any spring that is strained by more than $\epsilon$, and and 
bond angle that is changed by more than $\delta$ are removed from the network.
This choice was motivated by observations from recent mechanical studies 
which show that a trabecular architecture from a given location fails at a fixed level 
of strain, {\it independent of the strength or elastic modulus of the 
bone}~\cite{hogAruh}. However, bone samples from distinct locations exhibit
different levels of fragility. For example samples from 
proximal rat tibia, human tibia, and human lumbar spine have been shown to fracture 
at strain levels of approximately 5\%, 1\% and 7\% 
respectively~\cite{hogAruh,rohAlar,mosAmos}.

The first three networks represented in Figure 3 included a common fracture 
criterion; specifically $\epsilon = \epsilon_0 = 5\%$ and $\delta = \delta_0 = 0.1$. Since 
$\Gamma(\nu)$ is independent of $\epsilon$ and $\delta$ and 
$U(\nu)$ can be expected increase with them, $U(\Gamma)$ 
will depend on $\epsilon$ and $\delta$. The symbols `$\diamond$' in Figure 3
represent a fourth network~\cite{para3}, where the ultimate stress $U(\nu)$ has been rescaled
by a factor $(\epsilon_0 / \epsilon_1)$,  $\epsilon_1$ being the 
new value of the fracture strain. For parameters
chosen, failure of elastic elements (as opposed to bond-breaking) was the
dominant (though not exclusive) mode of fracture, justifying the use of
this rescaling factor~\cite{single}.
Once $U(\nu)$ is rescaled, points representing all networks 
collapse to the same bi-linear curve~\cite{error}.

The transition between the two linear segments of $U(\Gamma)$ is accompanied by 
a qualitative change in the stress distribution on compressed networks. 
Points on the right segment of each curve represent ``healthy" networks;
i.e., those with smaller values of $\nu$. The histograms of stresses for these networks,
an example of which is shown by the solid line of Figure 4, exhibit broad peaks.  
The presence of such peaks  is consistent with the geometry of their stress backbones, see Figure 1(a). In contrast, points on the left segment of $U(\Gamma)$ represent
``weak" networks (i.e., larger $\nu$), whose stress histograms exhibit no
broad peaks, and have exponential tails~\cite{chaAmac}. This behavior of the
histograms is consistent with the presence of a sparse stress backbones. 
The transition between the two linear segments of $U(\Gamma)$ with increasing $\nu$
coincides with the elimination of the peak in the histogram.

\begin{figure}
\epsfxsize=2.50truein
\hskip 0.25truein
\epsffile{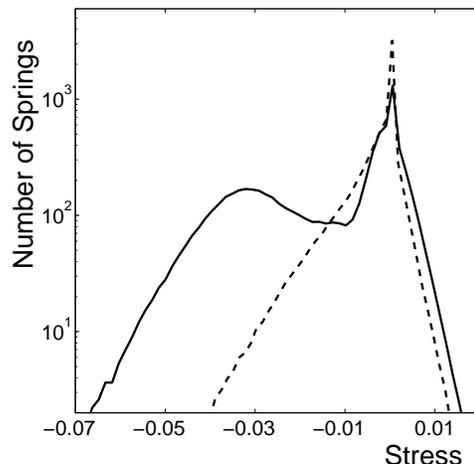}
\caption{The solid line shows a histogram of stresses for a compressed
network modeling a healthy TA ($\nu=10\%$). The dashed line shows the analog
for a weak TA ($\nu=25\%$). Each histogram represents the average
of 50 configurations with identical control parameters. The elimination of the 
broad peak of the histogram coincides with the transition between the two linear 
segments of $U(\Gamma)$. Positive and negative values of the stress denote 
extended and compressed springs respectively.}
\label{hist}
\end{figure}

The ultimate stress $U$ of a bone is the measure essential for
management of osteoporosis. Unfortunately, it is not accessible
in-vivo (without breaking a bone!), and surrogates such as 
the bone density are used to estimate $U$. 
The BMD of a patient is compared with that of a sample population
to determine if and when therapeutic interventions are necessary.
However $U$ is known to depend on other factors of bone such as the
architecture of its TA and the ``quality" of bone material. The significant  variations introduced 
by these factors makes it difficult to identify individuals susceptible
to fracture using measurements of BMD alone~\cite{marAjoh}.

These problems can be avoided if  a characteristic
that relates $U$ to an intrinsic property (i.e., one that does not need
to be compared to that of a population) of a bone is available. Since 
factors such as bone quality and architecture of the porous medium 
effect both  $U$ and $\Gamma$ it is conceivable 
that the relationship between them remains unchanged between patients.
Numerical analyses of our model system justify this expectation when $U$ is 
rescaled by a location dependent factor (quantifying the stiffness and fracture
strain of trabecular elements). 

Vibrational analysis has been used for in-vivo studies of bone
strength and protocols are available to obtain measurements needed
to evaluate $\Gamma$~\cite{couAhob,kauAein}. Once  the stiffness and 
fracture strain of different bone locations are 
tabulated, rescaling factors required for Figure 3 can be determined. 
Subsequently,  the ratio $\Gamma$ of responses of a bone to stationary and 
periodic strain can be used as a non-invasive diagnostic of bone strength.

The author would like to thank S. R. Nagel for pointing out that nonlinear 
response is related to the stress backbone.
He also acknowledges discussions with M. P. Marder  
and S. J. Wimalawansa. This research is partially funded by the National Science Foundation, 
the Office of Naval Research and the Texas Higher Education Coordinating
Board.

\end{multicols}

\begin{thebibliography}{99}

\bibitem{kibAsmi} P. Kiberstis, O. Smith, and C. Norman, 
   {\em Science}, {\bf 289}, 1497 (2000).

\bibitem{wein} R. S. Weinstein,  {\em J. Bone. Miner. Res.}, {\bf 15}, 621 (2000).

\bibitem{cann} C. E. Cann, {\em Radiology}, {\bf 140}, 813 (1981).

\bibitem{njeAhan} C. F. Njeh, D. Hans, T. Fuerst, C.-C Gl\"uer, and 
  H. K. Genant, ``Quantitative Ultrasound: Assessment of Osteoporosis and 
  Bone Status," Martin Dunitz, London (1999).

\bibitem{potAben} L. Pothuaud, C. L. Benhamou, P. Porion, E. Lespessailles, 
  R. Harba, and P. Levitz,  {\em J. Bone. Miner. Res.}, {\bf 15}, 691 (2000).

\bibitem{legAcha} E. Legrand, D. Chappard, C. Pascaretti, M. Duquenne, 
  S. Krebs, V. Rohmer, M-F. Basle, and M. Audran, {\em J. Bone. Miner. Res.},
  {\bf 15}, 13 (2000).

\bibitem{majAnew} S. Majumdar, D. Newitt, M. Jergas, A. Gies, E. Chiu, 
  D. Osman, J. Keltner, J. Keyak, and H. K. Genant, {\em Bone}, {\bf 17}, 417 (1995).

\bibitem{gunAbas} G. H. Gunaratne, K. E. Bassler, K. K. Mohanty, and 
  S. J. Wimalawansa, ``A Model for Bone Strength and Osteoporotic Fracture," 
  cond-mat/0009221 at http://xyz.lanl.gov. 

\bibitem{chuAroo} J. W. Chung, A. Roos, J. Th. M. De Hosson, and 
  E. Van der Giesses,  {\em Phys. Rev. B}, 
  {\bf 54}, 15094 (1996).

\bibitem{cowAsad} S. C. Cowin, A. M. Sadegh, and G. M. Luo, 
    {\em J. Biomech. Eng.}, {\bf 114}, 129-136 (1992).

\bibitem{para} These networks were obtained by randomly displacing vertices 
  by up to 1 unit from a square lattice
  with sides of 10 units. The elastic and bond-bending spring constants were 
  chosen within [0.5, 1.5] units  and [2.5, 7.5] units respectively,
  while $\epsilon=5\%$ and $\delta=0.1$. Each mass was 1 unit,
  and the dissipation coefficient $\eta=0.1$. Distinct networks with these 
  control parameters were generated by using different random seeds.

\bibitem{mouAdux} C. Moukarzel and P. M. Duxbury, Phys. Rev. Lett.,
  {\bf 75}, 4055 (1995).

\bibitem{foot2} The stationary states of the system are evaluated  
  using the conjugate gradient method to minimize potential energy.
  The dynamical properties of the network are calculated using the 
  Bulirsch-Stoer method~\cite{preAfla}.

\bibitem{fung} Y. C. Fung, ``Biomechanics: Mechanical Properties of Living 
  Tissue," Springer-Verlag, New York (1993).

\bibitem{foot3} Even though $\chi_{\nu} (t)$  and $T_{\nu} (t)$ can depend 
  on the driving frequency $\Omega_0$, numerical studies show no such 
  dependence when $\omega$ is scaled by $\Omega_0$.

\bibitem{preAfla} W. H. Press, B. P. Flannery, S. A. Teukolsky, and 
  W. T. Vettering, ``Numerical Recipes - The Art of Scientific Computing," 
  Cambridge University Press, Cambridge, 1988. 

\bibitem{para2} For this network, elastic and bond bending spring 
  constants were chosen within [0.3, 1.7] units and [0.4, 3.6] units 
  respectively,  while $\eta=0.001$ and $\zeta_0=5.0$ units. The exponential decay 
  of $U(\nu)$ was significantly faster for this network.

\bibitem{hogAruh} H. A. Hogan, S. P. Ruhmann, and H. W. Sampson,
J. Bone Miner. Res., {\bf 15}, 284 (2000);  C. M. Ford and T. M. Keaveny, 
J. Biomech., {\bf 29}, 1309 (1996).

\bibitem{rohAlar} L. Rohl, E. Larson, F. Linde, A. Orgaard, and J. Jorgensen,
J. Biomechanics, {\bf 24}, 1143 (1991).

\bibitem{mosAmos} L. Mosekilde, L. Mosekilde, and C. C. Danielsen, 
Bone, {\bf 8}, 79 (1987).

\bibitem{para3} The elastic and bond bending spring constants for this network
varied within ranges [0.5, 1.5] and [2.5, 7.5]  respectively, while $\eta=0.01$ and
$\zeta_0=3.0$. Fracture criteria are given by $\epsilon=2\%$ and $\delta=0.04$.

\bibitem{single} To the best of the author's knowledge, there has not been a study
of fracture of single trabeculae or bond-angles. The dominant mode of fracture 
(critical strain vs. bond breaking) of TAs from bone samples is also not known.

\bibitem{error} The standard error for the left and right segments were
  0.17 and 0.40 respectively. In the absence of an independent estimate of
  standard deviations, it is not meaningful to calculate the goodness of fit
  for the least square estimates. See Ref.~\cite{preAfla} for details.

\bibitem{chaAmac} S. Chan and J. Macha, Phys. Rev. B, {\bf 49}, 120 (1994).

\bibitem{marAjoh} D. Marshall, O. Johnell, and H. Wedel, 
  {\em BMJ},  {\bf 312}, 1254-1259 (1996); L. J. Melton, E. A. Chrischilles, 
  C. Cooper, A. N. Lane, and B. L. Riggs,  {\em J. Bone. Miner. Res.}, {\bf 7}, 1005 (1992).

\bibitem{couAhob} B. Couteau, M.-C. Hobatho, R. Darmana, J.-C. Bribnola, and J.-Y. Arlaud,
J. Biomech., {\bf 31}, 383 (1998); G. Van der Perre, R. Van Audekerke, M. Martens, and 
J.-C. Mulier, J. Biomech. Eng., {\bf 105}, 215 (1990).

\bibitem{kauAein} J. J. Kaufman and T. A. Einhorn, Osteoporos. Int., {\bf 8}, 517 (1993).

\end{thebibliography}
\end{document}